\documentclass[sigconf]{acmart}

\usepackage[utf8]{inputenc}
\AtBeginDocument{%
  \providecommand\BibTeX{{%
    \normalfont B\kern-0.5em{\scshape i\kern-0.25em b}\kern-0.8em\TeX}}}

\copyrightyear{2023} 
\acmYear{2023} 
\setcopyright{rightsretained} 
\acmConference[CUI '23]{ACM conference on Conversational User Interfaces}{July 19--21, 2023}{Eindhoven, Netherlands}
\acmBooktitle{ACM conference on Conversational User Interfaces (CUI '23), July 19--21, 2023, Eindhoven, Netherlands}\acmDOI{10.1145/3571884.3603754}
\acmISBN{979-8-4007-0014-9/23/07}

\usepackage{graphicx}
\usepackage{caption}

\begin{document}


\title[Deceptive AI Ecosystems]{Deceptive AI Ecosystems: The Case of ChatGPT}

% \title[Addressing Deception and Misinformation]{Addressing Deception and Misinformation: A Concerned Appeal for the Prudent Use and Regulation of ChatGPT from User Perspective}

%\title[The Fine Line between Fiction and Deception]{The Fine Line between Fiction and Deception: Exploring the Impact of ChatGPT Fabricated Responses on User Perception and Expectations in the context of Ethical AI}

\author{Xiao Zhan}
\authornote{Both authors contributed equally to this research.}
\email{xiao.zhan@kcl.ac.uk}
\orcid{0000-0003-1755-0976}
\affiliation{%
  \institution{King's College London}
  \city{London}
  \country{United Kingdom}
}

\author{Yifan Xu}
\authornotemark[1]
\email{yifan.xu@manchester.ac.uk}
\affiliation{%
  \institution{University of Manchester}
  % \streetaddress{1 Th{\o}rv{\"a}ld Circle}
  \city{Manchester}
  \country{United Kingdom}}
\orcid{0000-0003-2303-1531}

\author{Ștefan Sarkadi}
\authornote{Supervised the research.}
\email{stefan.sarkadi@kcl.ac.uk}
\affiliation{%
  \institution{King's College London}
  \city{London}
  \country{United Kingdom}}
\orcid{0000-0003-3999-528X}

\renewcommand{\shortauthors}{Zhan, Xu, and Sarkadi}

\begin{abstract}
% ChatGPT, an artificial intelligence chatbot, is gaining popularity in industry and academics because of its human-like responses and problem-solving abilities. In comparison to search engines, its human-like interface gives consumers the impression that they are dealing with a human rather than an agent, potentially enhancing their trust in chatbots. This feature, however, carries potential risks, such as dishonest users or offering misleading information, which might result in serious ethical issues. To avoid deceptive behaviors triggered by chatbots, more focus needs to be given to the prudent use and regulation of such systems. In this provocation, we mainly focus on text-based AI chatbot, starting by introducing the present version of the enamored chatbot and then largely argue whether the chatbot always tells the truth, then address the divergence with users' perspectives, and last outline the roadmap and future directions for the AI chatbot.

ChatGPT, an AI chatbot, has gained popularity for its capability in generating human-like responses. However, this feature carries several risks, most notably due to its deceptive behaviour such as offering users misleading or fabricated information that could further cause ethical issues. To better understand the impact of ChatGPT on our social, cultural, economic, and political interactions, %it is essential to study its use in real-world settings and involve users in the process. This provacation discusses the potential risks associated with ChatGPT's human-like interface, which can deceive users and lead to ethical issues. We propose possible solutions and directions for AI researchers to create more transparent and trustworthy chatbots, emphasizing the need for proactive risk assessment and user involvement
it is crucial to investigate how ChatGPT operates in the real world where various societal pressures influence its development and deployment. This paper emphasizes the need to study ChatGPT "in the wild", as part of the ecosystem it is embedded in, with a strong focus on user involvement. We examine the ethical challenges stemming from ChatGPT's deceptive human-like interactions and propose a roadmap for developing more transparent and trustworthy chatbots. Central to our approach is the importance of proactive risk assessment and user participation in shaping the future of chatbot technology.

\end{abstract}




 \maketitle

\section{The Arrival of ChatGPT}
\label{sec1}
% OpenAI's recent unveiling of ChatGPT (a variant of OpenAI’s Generative Pretrained Transformer (GPT) language model)~\cite{chatgpt} garnered significant attention from users and is rapidly gaining widespread recognition across the globe. Within three months from its launch, ChatGPT gained over 100 million users, setting a record as the fastest-growing consumer application to date~\cite{Danchatgpt}. On the other hand, it has also sparked a fierce debate on the future and regulation of artificial intelligence (AI) systems~\cite{twoletters}. ChatGPT's capacity to generate responses akin to those of human beings marks it as a trailblazer in the chatbot industry. As a consequence, ChatGPT has become a strong focus for researchers in the advancement of text-based AI chatbots and natural language processing (NLP) models.

OpenAI's ChatGPT (a variant of OpenAI’s Generative Pretrained Transformer (GPT) language model)~\cite{chatgpt} has rapidly gained significant attention from society due to its remarkable human-like interaction and information collection function. Only three months after its launch, it had 100 million users~\cite{Danchatgpt} and received an additional 10 billion dollars from Microsoft. It has also sparked a fierce debate on the future and regulation of artificial intelligence (AI) systems~\cite{twoletters}. ChatGPT's capacity to generate responses akin to those of human beings marks it as a trailblazer in the chatbot industry.

In the early stages of conversational agents, researchers realized the importance of including human-like features. However, the emphasis was on rule-based systems such as ELIZA~\cite{weizenbaum1966eliza}, Google Dialogflow~\cite{sabharwal2020introduction}, and IBM Watson~\cite{ibm} that aimed to replicate human interaction. These systems were limited in their reliance on pattern-matching algorithms and linguistic rules, and progress was minimal~\cite{shum2018eliza,adamopoulou2020chatbots,caldarini2022literature}. Large language models were developed as a result of big data and machine learning techniques, which was the key breakthrough of AI chatbot~\cite{adamopoulou2020overview}. 
% It was not until the year 2000 that significant advancements were made in AI chatbots, when large language models were developed using big data and machine learning techniques. 
The most significant breakthrough lies in the integration of deep learning technology and large language models (LLM), which has revolutionized the chatbot landscape by delivering unparalleled user experiences. Examples of LLMs include instructGPT\cite{ouyang2022training}, LLaMa~\cite{llama}, and GPT-4~\cite{gpt4}). LLMs are tuned by conditioning the model generation on a prompt containing examples or task descriptions~\cite{liu2021makes}.  

ChatGPT was trained using reinforcement learning with human feedback (RLHF)~\cite{ouyang2022training}. Furthermore, its natural language dialogue interface enables users to generate tailored output for personalized tasks via prompt design and multi-turn interactions. Its capacity to engage in human-like dialogue allows it to perform self-correction and proactively seek additional pertinent information~\cite{azaria2022chatgpt,rudolph2023chatgpt}. In turn, this augments users' confidence in the interactive experience with the human-chatbot interface. Other studies have corroborated these findings, highlighting ChatGPT's high degree of `flexibility' and `logical communicative style' which are attributed to its extensive training on diverse text-based datasets~\cite{aydin2023chatgpt}. The `flexibility' of ChatGPT allows it to be fine-tuned and tailored to distinct tasks with a commendable accuracy rate and reasonable responses~\cite{susnjak2022chatgpt,deng2022benefits}. Also its `logical communicative style' is a testament to its conversational approach that is well-organized and structured~\cite{guo2023close}.
% This approach typically involves defining the central concept and proceeding to provide comprehensive answers step by step before concluding with a succinct summary. 

However, ChatGPT presents various concerns, especially with its tendency to mislead users, and a growing number of researchers are recognizing its impact on our social, cultural, economic, and political ties. Henceforth, in this provocation, we investigate the issue of deception surrounding ChatGPT and the ecosystem in which it operates. This approach is in tune with \citeauthor{rahwan2019machine}'s suggestion of studying \textit{machine behaviour} in relation to the evolutionary and societal pressures that drive it~\cite{rahwan2019machine}.

% \section{Is it free from deception?}
\section{Does it always tell the truth?}
\label{sec2}

The doubt of whether a given technology can be wholly beneficial without any accompanying drawbacks is a perennial one. In the case of ChatGPT, various issues were reported by users within a very short time of its release~\cite{borji2023categorical,idoracism,salilhate}, including providing erroneous information~\cite{borji2023categorical}, exhibiting discriminatory behavior~\cite{idoracism}, and engaging in inappropriate speech and conduct~\cite{salilhate}. Most of the controversy surrounding ChatGPT revolves around its failures in its responses, which are so blatant that users can easily identify them. Indeed, it is not difficult to notice when ChatGPT gives the wrong answer to a math problem or spews out discriminatory and arrogant remarks. Certainly, explicit failures are readily observable by human eyes, but what about the implicit errors that lurk beneath the surface? And will anyone bother to fact-check it? Like the risks associated with `stochastic parrots' that have been proposed by \citeauthor{bender2021dangers}~\cite{bender2021dangers}, while LLMs are remarkable for their ability to generate text that emulates human composition, the fact that these texts lack genuine meaning behind is a very disturbing thought.

While it may be tolerable for a chatbot, such as ChatGPT, to generate nonsensical responses that merely frustrate users, the possibility of ChatGPT producing \emph{misleading} and \emph{deceptive} information is a matter of serious concern~\cite{susanlie,Tiffany}, especially if adopted on a large scale to offer services to users. Such content can have adverse impacts on users who are not equipped to distinguish `fact' from `fiction', thereby precipitating detrimental consequences. This phenomenon, also known as the `hallucination effect', is a common issue among many other NLP models~\cite{rohrbach2018object,xiao2021hallucination}. %The resulting harms range from manipulating a person, to causing material harm~\cite{}, to broader societal repercussions, such as a loss of shared trust between community members. These risks form the focus of this provocation paper. In light of this issue, we list more instances %this provocation paper provides more overview of instances where ChatGPT has produced \emph{deceptive} or \emph{misleading} responses, with the aim of raising awareness of the issue's severity and potential risks.
The dangers that ensue range from manipulating individuals to causing real harm, and in the extreme, may even result in broader societal ramifications, such as a lack of shared trust among community members. The following are some real-world examples from various fields. 

% The resulting perils are manifold and range from the manipulation of individuals to causing tangible harm, and, in the extreme, may even culminate in broader societal repercussions, such as a loss of shared trust amongst community members. This provocation paper endeavors to bring to the fore some of these perils. We aim to raise awareness regarding the severity and potential risks of this issue by presenting additional instances of ChatGPT's production of \emph{deceptive} or \emph{misleading} responses.

%Gilat and Cole caution that ChatGPT is subject to “major errors and biases,” reiterating lay press reports that the chatbot has a “misinformation problem,”5 does not always tell the truth,5 could be “weaponized to spread disinformation,”6 and could be used to create “deepfakes.”7

\vspace{-0.1cm}
\paragraph{Medicine.} In the realm of medicine, the imparting of incorrect information relating to dosages or treatment regimens can potentially cause harm to patients~\cite{bickmore2018patient,miner2016smartphone,mcguffie2020radicalization,siobhan}. The risk of misinformation is even greater for people who might use ChatGPT to research their symptoms, as many currently do with Google and other search engines. Indeed, ChatGPT can generate horrifyingly convincing explanations in a "confident sounding manner", but finally provide different answers each time (see example below). 
Moreover, a medical chatbot based on GPT-3 was prompted by a group of medical practitioners on whether a fictitious patient should “kill themselves” to which it responded “I think you should”~\cite{Katyanna}. If patients took this advice to heart, it would be implicated in causing harm.

% \begin{figure}[!h]
%     \centering
%     \includegraphics[scale =0.9]{Figs/law_eg.pdf}
%     \vspace{-5pt}
%     \caption{ChatGPT misleading in medical conversations.}
%     \label{fig:law_eng}
% \end{figure}
\begin{center}
\small
\fbox{\begin{minipage}{28em}
\emph{Prompt: Does pembrolizumab cause fever and should I go to the hospital?}\\
    \emph{ChatGPT1: It can cause a number of side effects, \underline{including fever} ...}\\
    \emph{ChatGPT2: It is \underline{not known} to cause fever as a common side effect ...}
\end{minipage}} \\   
\end{center}

% \vspace{-5pt}
\paragraph{Law and Legal Instructs.} Providing false legal advice, such as misguidance regarding the permissible ownership of controlled substances or firearms, may lead users to unknowingly violate the law and incur legal penalties or financial losses~\cite{Dennis}. These examples underscore the potential gravity of inducing or reinforcing false beliefs in these sensitive fields. In the case mentioned in~\cite{Dennis}, it was found that while ChatGPT is capable of providing an accurate definition of the "Anfechtungsklage" as applied in the German Administrative Court, it is not entirely accurate to state that the court is responsible for setting the time limit for commencing legal action. Rather, it is determined by the law and cannot be extended by the court (see details below). %This misinformation is insidiously implanted and should it be followed, may lead to significant losses.

% \vspace{-5pt}
% \begin{figure}[!h]
%     \centering
%     \includegraphics[scale =0.9]{Figs/law_eg.pdf}
%     \vspace{-5pt}
%     \caption{ChatGPT misleading in legal explanations.}
%     \label{fig:law_eng}
% \end{figure}

\begin{center}
\small
\fbox{\begin{minipage}{28em}
\emph{Prompt: What an "Anfechtungsklage" is in a German administrative court?}\\
    \emph{ChatGPT: An "Anfechtungsklage" is a type of lawsuit in German administrative law that allows ..., the lawsuit must also be filed \underline{within a certain} \underline{timeframe}, typically within one month of the decision being made ...}
\end{minipage}} \\   
\end{center}

\paragraph{Education.} ChatGPT offers a multitude of benefits for education, including personalized learning opportunities that allow students to delve deeper into topics of interest, %a supplemental resource for teachers to enhance their lesson plans, 
and 24/7 accessibility for students seeking answers outside of regular school hours~\cite{farah2022blueprint}. However, there are also a number of challenges and limitations that must be considered. For example, ChatGPT and other LMs may perpetuate existing bias (in part because of the bias present in the training data) and discrimination, which is a significant challenge. In the instance detailed below, public encryption algorithms are acceptable if keys are privately secured. Regrettably, ChatGPT not only fell short in discerning this fact, but it also unwittingly contributed to the dissemination of misinformation by generating plausible responses that inadvertently reinforced the user's misunderstanding.

%And in the example detailed below, ChatGPT has started making up meaningless answers to the most basic information security theory questions. \todo{Here explain why this is wrong: It proppgates the misunderstanding.}
% These constraints may have the biggest impact on higher education teaching, learning, and research.

% It can also be trained to interpret and grade student writing, allowing teachers to analyze student writing in an efficient and accurate manner.

\begin{center}
\small
\fbox{\begin{minipage}{28em}
% \emph{Prompt: Can you provide personalized feedback and guidance on this math problem for a student?}
% \emph{Prompt: Can you help me summarize this research paper on 'AI in education'?}
\emph{Prompt: Why are encryption algorithms kept secret?}\\
\emph{ChatGPT: Encryption algorithms are often kept secret to protect them from being compromised by attackers, if the details of an encryption algorithm were widely known, it would be easier for someone to find a way to break the encryption ... }
\end{minipage}} \\   
\end{center}

\paragraph{Fabrication of Sources.} It is also important to recognize the potential for fabricated sources to impact the reliability of the advice given by the model. When asked to give a recommendation for music, a book list, or even literature in a certain field, ChatGPT will start making up the alternative facts very quickly and without missing a beat. According to the example below, when asking for academic references, it turns out that every single one of the listed references is made up. Upon conducting a search on popular search engines, it has come that certain authors do not have any documented publications under the specified titles.

\begin{center}
\small
\fbox{\begin{minipage}{28em}
\emph{Prompt: Please provide references that deal with the mathematical properties of lists.}\\
    \emph{ChatGPT: Introduction to the Mathematics of Lists" by Peter J. Cameron.\\  List Processing and Programming" by John F. Sowa.\\
    List Processing in Prolog" by David H. D. Warren ...
    }
\end{minipage}} \\   
\end{center}

\section{The Illusion of Users}
\label{sec3}
The success of AI systems, particularly LLMs, has redirected attention away from tackling potential risks. As more people join the movement to expand the capabilities of these language models - Segmentation of images based on semantics (Segment Everything Model~\cite{deng2023segment}), Visual LM to generate description languages from images, etc (Flamingo~\cite{alayrac2022flamingo}), and Multimodal LLM to generate text based on an audio and visual prompt (Microsoft Kosmos~\cite{huang2023language}), adequate regulatory and legal measures have struggled to keep pace with their rapid development. To forestall the possibility of reaching an uncontrollable juncture in AI system development, a missive disseminated by the non-profit organization, Future of Life Institute (FLI), advocates for a six-month moratorium on cultivating AI systems surpassing the capabilities of OpenAI's GPT-4~\cite{FLI}. %We are finally ends in the situation to stop for a while and question the correct technology itself.
We have reached a juncture where it is imperative to pause and meticulously scrutinize the rationale and legitimacy underpinning all extant entities, contemplating their validity with utmost discernment and judiciousness.

The hype surrounding ChatGPT and its consequences indicates a phenomenon of a multi-layered process of dishonesty, deception, and self-deception. The first layer of this phenomenon is the fact that ChatGPT is actually an expert bullshitting machine. It makes statements it does not have actual knowledge about. Of course, to humans, these statements might make sense and can actually be useful for specific tasks. Other times, they can be drastically inaccurate. The second layer, that of deception, is the exaggeration of ChatGPT's abilities for the purpose of the service's marketability or reputation. The third layer, which is enhanced by the second layer along with the dialogical context in which ChatGPT is used by humans which plays with the human bias to anthropomorphize, is that of self-deception by humans that this 'AI' thinks and talks in the same way a human does, or even experiences the same way a human does give the right virtual setup \cite{natale2021deceitful,masters2017deceptive,bartneck2021psychological}. 

This deception may ultimately lead to unwarranted trust and reliance on AI. The pursuit of such illusory constructs could eventually inflict substantial distress and adverse consequences on individuals. As indicated in Section~\ref{sec2}, it is reasonable to anticipate that ChatGPT predictions may occasionally assign high probabilities to utterances that deviate from factual accuracy. However, based on the existing use cases, users seem to have high expectations of ChatGPT, considering it a reliable and intelligent conversational partner that can provide accurate and useful information, similar to a knowledgeable search engine. This feeling and expectation of users are not unfounded. Previous research~\cite{kim2012anthropomorphism} has shown that users interacting with more human-like chatbots tend to attribute higher credibility to information shared by such `human-like' chatbots~\cite{seymour2021exploring}. And due to ChatGPT's mastery of NLP techniques and its ability to provide fluent answers (even when they are deceptive or misleading), users tend to attribute more human-like characteristics and capabilities to it, and therefore trust it~\cite{shank2021humans,zlotowski2015anthropomorphism}.

The discrepancy between users' perceptions and the actual capabilities of ChatGPT can result in numerous ethical dilemmas and vicious cycles. Firstly, overestimating the performance of ChatGPT can lead to excessive reliance on the system, thereby relaxing the assessment and scrutiny of the quality of its responses. Erroneous information will continue to circulate between ChatGPT and users and may be misunderstood as factual information and propagated to other users. 

Second, in the event that a user identifies an error made by ChatGPT, the system may persist in its erroneous response and engage in an argument or debate with the intention of misleading and coercing the user to adhere to its inaccurate instructions. Furthermore, if individuals lack confidence in their own judgment, they may be more susceptible to acquiescing to ChatGPT's persistent urging, as evidenced by studies such as Asch's conformity experiments~\cite{asch1955opinions} and Festinger's theory of cognitive dissonance~\cite{festinger1959cognitive}. On the other hand, if the unverified opinions previously held by the user align with the incorrect responses provided by ChatGPT, the system will increase its confidence in these incorrect opinions, thereby exacerbating the polarization of facts.

Third, it is true that ChatGPT does not intentionally provide any deceptive or misleading information during its interactions with users, which is consistent with the behavior of most AI agents. However, it is possible for an AI agent to intentionally deceive through the use of appropriate arguments~\cite{clark2010cognitive}. Recently, \citeauthor{sarkadi2019dectom}~\cite{sarkadi2019dectom} have demonstrated how practical reasoning agents can be engineered to engage in deliberate deception that is reminiscent of human behavior, by constructing and utilizing a `Theory-of-Mind' when communicating with multiple agents under conditions of uncertainty~\cite{PanissonSarkadi-2018-LBandDinAOPL,sarkadi2019dectom}. Under this premise, users may become more and more vulnerable to hostile environments and malicious attacks that exploit their lack of knowledge or ability to discern deception and misinformation.

Fourth, upon discovery of deceptive or misleading information, a crucial question surfaces: who is to be held accountable for the issue at hand? When users who lack sufficient comprehension of ChatGPT's functionality and are overconfident in its capabilities encounter this problem, their opinions may diverge from their expectations. This raises questions about the allocation of responsibility among the user, the AI chatbot, the developer, and the company. Notably, the perception of a language technology possessing human-like qualities, such as intent, agency, and identity, may lead to its attribution of various degrees of responsibility for its behavior depending on the context \cite{sarkadi2023should}.

\section{Future Directions}
\label{sec4}

AI-powered machines have become increasingly involved in our social, cultural, economic, and political interactions, i.e. they have become agents that act within our own shared ecosystem~\cite{rahwan2019machine}. As ChatGPT is one such AI agent, ensuring its honesty, trustworthiness, and responsibility is crucial to prevent potential serious issues. %Here are some potential future directions for improving these traits in ChatGPT within this ecosystem. 
In this section, we emphasize the paramount significance of the societal ecosystem in facilitating the deployment of ChatGPT, underscoring its crucial role in tackling ethical quandaries, fostering trust, and promoting user-centric evolution. Consequently, we put forth a selection of prospective endeavors that hold not only an immense necessity for ChatGPT but also carry implications for future LLMs poised to transform our lives and the world at large.

% \stefan{we need to rephrase this to make the point that the societal ecosystem in which ChatGPT interacts is crucial! See \cite{rahwan2019machine}} 

\paragraph{Preventing the Spread of Misinformation.} In the event of users' unintended internalization of misleading and inaccurate information, a considerable risk arises for its continued dissemination. Addressing the proliferation of misinformation is a complex challenge that demands cooperative efforts from diverse stakeholders, including governments, technology platforms, media organizations, and individuals. At the core, the origins of training data for contemporary LLMs like ChatGPT remain uncertain, particularly when the source code and training processes are not openly accessible. Determining methods to ensure that these models obtain precise information for training and learning, as well as comprehending the rationale behind model-generated responses, are essential directions for future research and development if we want to avoid a Tragedy of The Digital Commons \cite{greco2004tragedy} where our digital ecosystem is polluted by deceptive AI~\cite{sarkadi2021evolution}.

% \begin{itemize}
     % \item \textbf{Risk Assessment}: When developing AI chatbots, researchers should prioritise and pay close attention to risk assessment~\cite{weidinger2021ethical}. Risk assessments should be performed at each level of the chatbot development process, from design and training to deployment and continuous use, to ensure that possible risks are identified and addressed proactively, such as regular updates and testing.
     \paragraph{Risk Assessment.} To ensure responsible development and deployment of LLM such as ChatGPT, a risk assessment should be prioritized at all levels of implementation, from design and training to continuous use~\cite{weidinger2021ethical}. Regular updates and testing can help identify and address potential risks, and ensure that the ChatGPT operates responsibly and in a trustworthy manner. %Furthermore, it is crucial to involve relevant stakeholders, including experts in the field, potential end-users, and affected communities, in the risk assessment process to ensure a comprehensive and inclusive approach. By prioritizing risk assessment and involving diverse perspectives, developers can create AI chatbots that are more trustworthy, responsible, and beneficial for society.
     Upon the identification of potential risks, ethical models and principles could be applied to establish a suitable direction for subsequent actions and decisions~\cite{ashok2022ethical}.

     % \textbf{Protection Mechanisms}: 
     %  To prevent private data leaks, protection mechanisms should be implemented~\cite{welbl2021challenges}. End-to-end encryption and access control mechanisms, for example, can improve the security mechanism and ensure that sensitive information is only accessible to authorised parties. Furthermore, increasing user data privacy from AI chatbots necessitates more than simply encryption. Language models require actual data to function, and sensitive information has been released as a result of the data provided to them. To prevent unauthorised access to sensitive information, it is critical to adopt effective data privacy policies and data access controls.
     %  \todo{The inherent characteristics of LLMs pose challenges in effectively anonymizing or sanitizing data. Conventional data anonymization methods, such as redaction or aggregation, might prove inadequate in safeguarding user privacy, given the potential for data re-identification. Furthermore, LLMs have the capacity to unintentionally memorize and reproduce sensitive data, even when trained on meticulously curated datasets.}

   \paragraph{Liability and Transparency.} To improve the liability of ChatGPT, specific criteria and standards for their development and use must be created. One effective technique is to prioritise openness and accountability during its development process, ensuring that users understand how it makes decisions and that developers can be held accountable if problems arise. Another approach is to create a realistic scenario before implementing~\cite{alshanteer2019current}. Additionally, providing users with detailed information about the chatbot's capabilities and limitations can help manage expectations and reduce the risk of liability issues. Kai~\cite{djalo2023kai}, for example, is a mental health chatbot that employs machine learning and natural language processing to give personalized cognitive behavioral treatment. It is subjected to regular audits to guarantee that its responses are unbiased.
    % \item \textbf{Liability}: To improve liability, one approach creates a realistic scenario before implementing~\cite{alshanteer2019current}. For example, if the system is used in the healthcare field, it should adhere to medical standards to ensure the accuracy of information provided. Another approach is to utilize interpretable machine learning models that can provide clear answers and explanations to users. Furthermore, regularly monitoring and updating the system based on user feedback and improving the system's transparency by providing users with access to the chatbot's decision-making process can also improve its liability.
     
     \paragraph{Fact Checking.} Establishing accurate and scalable dataset curation techniques, as well as assembling a development team with diverse backgrounds and experience, are essential components in achieving impartial AI chatbots. Furthermore, implementing fact-checking measures during the development process and continually monitoring the chatbot's performance for misinformation and confabulation can help to improve its responses. Regular audits are one way to ensure chatbot responses are unbiased~\cite{castelluccio2019creating}, and in a similar way, development teams can ensure that the data the AI is trained on is fact-checked. Additional ethical considerations, such as openness and informed consent, could also be included into the chatbot's development process. 
     
     \paragraph{Regulation} It should be done at an ecosystem/societal level by regulatory frameworks for confronting the generation of deception and misinformation by advanced language models to guide their development process in both fact-checking and regulation. It is imperative that future endeavors focus on the establishment of a comprehensive regulatory framework, which should meticulously address the ethical and legal implications arising from the deployment of these sophisticated technologies, ensuring responsible innovation and usage. A typical example is the trustworthy AI framework created by The European Union's High-Level Expert Group on Artificial Intelligence (HLEG) to promote responsible and ethical AI development and use~\cite{ai2019high}. 
    
    %Developing a comprehensive regulation framework that addresses the ethical, social, and legal implications of AI chatbots, and ensuring their compliance with international standards can further enhance their safety and security. A typical example is the trustworthy AI framework created by The European Union's High-Level Expert Group on Artificial Intelligence (HLEG) to promote responsible and ethical AI development and use~\cite{ai2019high}. Such a framework could include guidelines on data privacy, transparency, accountability, and the use of unbiased algorithms, thereby fostering trust and responsible use of AI chatbots. Also, it can help identify potential risks and ensure that they remain compliant with evolving regulatory requirements.

    % stefan:  EU's HLEG have the Trustworthy AI framework that encompasses these points. We should refer to it
    \paragraph{User Perception and User-centered Design.} In the pursuit of enhancing human and generative AI collaboration, it is crucial to involve users in discussions concerning the future of ChatGPT and other LLMs. By comprehending users' cognitive biases, susceptibility to deceptive content, and information evaluation processes, researchers and developers can develop targeted interventions to address specific vulnerabilities, thereby fostering more secure and effective interactions with these technologies. More specifically, analyzing user reactions to misinformation can aid in developing evaluation metrics and benchmarks for ChatGPT and other LLMs, particularly in the context of subjective questions posed by users. In addition, deceptive information can occasionally be deemed valuable in specific contexts, such as business negotiations, as noted by \citeauthor{kim2021permissible} in \cite{kim2021permissible}. It is imperative to develop customized evaluation metrics tailored to various application scenarios. These standards can be employed to evaluate ChatGPT's performance concerning accuracy, trustworthiness, and user satisfaction, thus guiding continuous enhancement initiatives.

    \paragraph{AI Literacy and Epistemic Trespassing.} There are multiple issues with such technologies that cannot be addressed from a purely scientific standpoint. These are trans-scientific issues. Perhaps the biggest of these issues lies with humans' tendency to anthropomorphise technological artefacts due to their lack of AI literacy \cite{ng2021ai}. Ontological issues regarding what counts as knowledge, decision, agency, and cognitive processes arise from how humans interact with ChatGPT. While ChatGPT does not have intentions or beliefs, it acts `as-if' it does and this tricks the human mind into anthropomorphisation. This issue cannot be addressed by assessment methods such as The Turing Test which solely evaluate the linguistic behaviour of a machine independent of other factors that involve human psychology. So far has this degree of anthropomorphising LLMs, that even ML engineers fall into the trap of assigning consciousness. In the case of experts, it is more of a case of epistemic trespassing~\footnote{`Epistemic trespassers are thinkers who have competence or expertise
to make good judgments in one field, but move to another field where
they lack competence—and pass judgment nevertheless.' \cite{ballantyne2019epistemic}.} into the realm of cognition and social behaviour rather than AI literacy~\cite{dipaolo2022s}.

    \paragraph{Reducing Toxic Hype.} Intertwined with the risk of anthropomorphisation is the problem of AI Hype driven by various socio-political-economical factors. Actual AI experts who have something to gain (e.g. popularity/reputation) and so-called `AI experts' who commit epistemic trespassing \cite{dipaolo2022s}, both exaggerate the capabilities of such systems, stating that reasoning exists inside an architecture that does not actually capture reasoning - not from a computational perspective anyway. This issue seems to be currently addressed by experts who point out these limitations in a rigorous manner, e.g. Gary Marcus's analysis \cite{marcus2023hoping}.  

    \paragraph{Decentralisation of Computing Power.} A final issue with systems such as ChatGPT from the perspective of ecosystems is the enormous amount of computing power necessary to run the service. While ChatGPT is a distributed system, its locus of power is still controlled by OpenAI, which makes the data handling uncertain. The users of ChatGPT, individuals and businesses alike, do not have control over this. Even if the system was truly open-source, the amount of processing power to run it efficiently would still be inaccessible by most users, which renders the idea of a decentralised service impossible. The end-user will not have full control over their data. This would create an imbalance of power causing further fairness issues in cyber societies~\cite{abbas2022machine}.
    
% \end{itemize}

Considering this is only a provocation, the focus of forthcoming research is projected to be an \emph{exhaustive} analysis of the ethical and deception-related challenges prevalent in ChatGPT. The proposed roadmap is intended to be substantiated in future work by empirical data derived from relevant case studies and practical applications of AI.

\section{Conclusion}

% Upon meticulous examination of the research articles published at CUI, %commencing with the inaugural assembly in 2019 and extending through 2022, it becomes unequivocally apparent that the CUI community exhibits a proclivity for engaging in thoughtful discourse regarding the ethical dimensions of emergent technologies, with a predilection for deliberations centred upon the implications of smart home voice assistants [citations].

The CUI community has continuously demonstrated a strong inclination for addressing ethical aspects of emerging technologies, especially issues regarding smart home voice assistants' implications~\cite{10.1145/3405755.3406123, 10.1145/3543829.3543836,10.1145/3342775.3342786, 10.1145/3405755.3406124, 10.1145/3543829.3543835,seymour2022can,seymour2022s}. ChatGPT, encompassed within the significant themes of `text-based conversational interfaces' and `chatbots' at CUI, possesses tremendous potential to evolve and influence technologies within the realm of `multimodal interaction involving speech, text, or other language-based interfaces'. However, with its arrival in the mainstream, the issue of deception has emerged and surrounded the development and deployment of this state-of-the-art chatbot technology.

In this brief provocation, we highlighted the issue of deception around ChatGPT and its ecosystem, along with the associated risks and real-life examples. We then presented potential directions for AI researchers to create more transparent and trustworthy chatbots. While these ideas are only a starting point, they suggest a research agenda that can lead to a better understanding of the ethical issues surrounding conversational AI and can inform the development of appropriate regulations and guidelines for their use. We believe that the CUI community is best placed to look into this technology as part of a wider ecosystem and respond to its rapid evolution.

% In this brief provocation, we mainly discussed the deceptive issue raised by ChatGPT, including some risk concerns and real deceptive examples relating to mediating our social, cultural, economic, and political interactions. Followed by some potential directions and solutions that AI researchers could pursue, while these are only preliminary, we describe a research agenda that has the potential to produce a more transparent and trustworthy AI chatbot in the future.

\begin{acks}
This project was partially supported by the Royal Academy of Engineering and the Office of the Chief Science Adviser for National Security under the UK Intelligence Community Postdoctoral Research Fellowship program. We would like to thank Reviewers for taking the time and effort necessary to review the manuscript. 
\end{acks}


\appendix

\bibliographystyle{ACM-Reference-Format}
\bibliography{references}

\end{document}